\documentclass[10pt,a4paper]{article}

\renewcommand{\author}{Emil Khalisi}
\newcommand{\titel}{Clusters of Solar Eclipses in the Maori Era}
\newcommand{\version}{Version 1.73}
\renewcommand{\date}{\today}

\usepackage{hyperref}
\hypersetup{pdfauthor={\author }}
\hypersetup{pdftitle={\titel }}
\hypersetup{pdfsubject={Online publication, \version\ of \date\ }}
\hypersetup{pdfkeywords={Eclipses, New Zealand, Maori, mythology,
    astronomical dating, Wellington, Polynesia.}}

\usepackage[T1]{fontenc}        
\usepackage{mathptmx}           
\usepackage{pifont}             
\usepackage[scaled=0.92]{helvet} 

\usepackage[british]{babel}     
\usepackage{amsmath}            
\usepackage{microtype}          
\usepackage{arydshln}           

\usepackage{calc}
\usepackage[a4paper]{geometry}  
\usepackage{scrextend}           
\changefontsizes{10pt}

\usepackage{titlesec}
\titleformat*{\section}{\large\bfseries}
\titleformat*{\subsection}{\normalsize\bfseries}

\usepackage{graphicx}           
\usepackage{float}              
\usepackage{caption}            
\usepackage{fancyhdr}
\renewcommand{\headrulewidth}{0.4pt}
\usepackage{colortbl}             
\definecolor{grey20}{RGB}{208,208,208}
\definecolor{grey10}{RGB}{230,230,230}

\begin{document}


\fancyhead{}
\fancyhead[LO]{%
   \footnotesize \textsc{In original form published in:}\\
   {\footnotesize Habilitation at the University of Heidelberg }
}
\fancyhead[RO]{
   \footnotesize {\tt arXiv: 2009.01663 [physics.hist-ph]}\\
   \footnotesize {v2: 6th September 2020}%
}
\fancyfoot[C]{\thepage}

\renewcommand{\abstractname}{}

\twocolumn[
\begin{@twocolumnfalse}

\section*{\centerline{\LARGE \titel }}

\begin{center}
{\author \\}
\textit{D--69126 Heidelberg, Germany}\\
\textit{e-mail:} \texttt{ekhalisi[at]khalisi[dot]com}\\
%
\end{center}


\vspace{-\baselineskip}
\begin{abstract}
\changefontsizes{10pt}
\noindent
\textbf{Abstract.}
A dozen of high-magnitude solar eclipses accumulated near New
Zealand in the 15th century AD when the Maori inhabited the two
main islands.
Taking today's capital Wellington as the point of reference, we
counted ten events with magnitude larger than 0.9 between 1409 and
1516 AD and two more just below this value.
The eclipses need not have been all observed on account of weather
conditions.
An allusion to a particular event that could be conveyed in a myth
is discussed, but the dating turns out far from certain.
We take the opportunity here to meet the astronomy of the Maori and
their understanding of this natural phenomenon.
Moreover, an announcement is made to a cluster of five central
eclipses of the sun that will encounter New Zealand from 2035 to
2045.

\vspace{\baselineskip}
\noindent
\textbf{Keywords:}
Eclipses,
New Zealand,
Maori,
mythology,
astronomical dating.
\end{abstract}

\centerline{\rule{0.8\textwidth}{0.4pt}}
\vspace{2\baselineskip}

\end{@twocolumnfalse}
]




\section{Introduction}

Every planet in our Solar System except Mercury and Venus features
eclipses.
Some of the extraterrestrial obscurations will appear unusual and
strange to us as regards totality and duration and, in some few
cases, their progress.
But they are nowhere as exciting as on Earth.

The variety of our terrestrial eclipses is owed to the fine-tuning
of several astronomical factors:
the apparent sizes of the sun and moon, the inclination of two
rotational axes against the orbital plane, and the non-negligible
tidal effect responsible for the slowing down of the daily rotation.
Also, the shapes of tracks exhibit a diversity being unique in the
Solar System.
Different to other planets, an observer on Earth may experience a
total or annular eclipse, compute certain cycles, and discover
special features for his geographical spot.
Both astronomy and mathematics push the door open to a bouquet of
ideas for investigation.
Involving historical aspects, the researcher is confronted with
another realm full of anecdotes and stories of life facing
ethnology, cultural history, and, most of all, scientific dating.
Several issues have been encountered in our previous papers
published on {\tt arXiv} within the course of the current year.

This paper focuses on the 14th and 15th century AD when an
extraordinary series of eclipses occurred in or near New Zealand.
Our list of events reveals four decades, at least, with three or
four closely spaced high-magnitude events.
The islands were inhabited by the Maori people at that time.
There is no straight account on eclipses known, but information on
the Maori past is often interwoven with mythology
in chants, tales, and folklore instead of writings.
Such practice differs significantly from civilisations bequeathing
persistent documents.
In this paper we seize the opportunity to introduce a little bit
of history of this people from the Pacific.


\section{Colonisation of New Zealand}

New Zealand remained untouched by man as one of the last spots
on the map.
The first pre-historic settlements emanated presumably from the
Polynesian Islands (Fiji, Tonga, and Samoa) at about 800--1000 AD
or, according to other estimations, at 200 BCE.
The devastating decline of the biota began shortly thereafter.
The ecological consequences upon the arrival of humans show
similarities to many other sites in terms of deforestation and
extinction of species.

There are two archaeological models that set the initial colonisation
to either $\approx$800 AD with a small founding population of 10
to 20 individuals, or a later colonisation at 1280--1300 AD with a
larger group of 100 to 200 individuals.
Both assumptions can be however reconciled by a pause of about 500
years between the immigrations.
Radiocarbon analysis supports the transformation of the wildlife
by the intrusion of the omnivorous Pacific rat from 1280 AD on
\cite{wilmshurst-etal_2008}.
This moment marks the earliest ``visibility'' of human presence.
Tradition constructs legends how the first settlers arrived to New
Zealand at a time as early as the 11th and 12th century
\cite{simmonds_2018}.
Today, an iconic date of 1350 AD is adopted and entrenched in the
nationwide education system.
It was established by Percy Smith (1840--1922), a New Zealand
ethnographer, upon a mere exercise in averaging out genealogical
lines in old narratives.

The first Polynesian settlers developed a distinct culture on
the islands, now known as Maori, and called their homeland
``Aotearoa'', often translated as ``land of the white cloud''.
Contact to European explorers opened up in 1642 when the seafarer
Abel Tasman (1603--1659) discovered the land, and James Cook
(1728--1779) came across a hundred years later mapping the
coastline.
Just as everywhere in the world, the colonial era faced
deconstruction of the cultural heritage of the Maori, displacement
of people, and disregard of their belief.
Our modern knowledge about the indigenous people is quite sketchy,
because useful relicts are rare.
It turns out difficult to grasp the pristine concepts of their
views on nature and creation.
A few petroglyphs are known but they depict predominantly humans
and animals, barely natural phenomena.
It will be daring to draw parallels between geometric figures and
celestial objects to interpret a possible cosmology.
Some ideas are fathomed out indirectly from the language itself.
We know, e.g., that the Maori gave proper names to celestial objects.
Besides planets and bright stars, also transient appearances like
comets and meteors got their own name.

Another issue is that the first ethnological studies were performed
in the late 19th century, more than 100 years after the contact
with the European explorers.
It is unclear how many and which details of the original thoughts
amalgamated with modern knowledge.
Especially the voyage by James Cook was undertaken under the
auspices of the Venus transit in 1769 involving astronomers
\cite{finsternisbuch}.
None of the scientists thought of retaining the native culture for
the record but rather proudly explained their own astronomical
activities presenting the latest optical devices.
A recent revitalisation of the pre-colonial understanding of the
sky was launched by Pauline Harris and Rangi Matamua among others
\cite{harris_2013}.


\section{Education in Astronomy}

The Maori people practised astronomy empirically.
It was entangled into a kind of astrology in their everyday life.
The astronomers were looked upon as ``weather prophets'' and had
the status of reliable advisers for travellers and fishermen
\cite{best_1922}.
They knew the movement of stars, watched for special manifestations
in the sky and made use of instantaneous appearances to forecast
atmospheric conditions.
From this they were able to gauge sea voyages, food-quest
activities, and other seasonal effects.

Astronomical knowledge was firmly protected inside a small circle
of adepts like priests, medicine men, chieftains, and persons of
rank.
The Maori operated astronomical schools visited by this circle
\cite{white_1887}.
Each year they assembled to compare their observations of the
heavenly bodies and discussed the relevance of omens.
Useful knowledge was exchanged about crops, fishing, or food
gathering.
Almost every village had such a school, according to the number of
its inhabitants.
The course lasted for about 3--5 months, and for non-members it was
strictly prohibited to even sojourn the proximity of the educational
lodge.
When at a short distance the passers-by had to call to those within
the hut.
However, a Maori individual of those days knew about the sky more
than an average person in our modern time.

Storytelling held a central place in Maori life.
Considered as an important skill, this was the way how history was
passed on, person to person down the generations.
Historical events were embedded cryptically in descriptions that
have become legends over time.
Many stories and myths of the Polynesians contain
``semi-historical'' realism, since they preserved remnant
information on important incidents as well as the natural
environment.
This regards the genealogical lineage of royal chiefs, volcanoes,
cyclonic storms, or local customs.
They are all embedded into the deeds of culture heroes who are
themselves based on real persons whose history has been transmuted
in great mythological cycles \cite{taonui_2006}.
Certain details can be found across the breadth of the Pacific
islands suggesting a common origin from about 2000 BCE.

Though recitation played a central role in Maori tradition,
education to an astronomer still must have been a long-winded task
without written records.
In general, word-of-mouth transfer did hardly suffice to acquire
the necessary practical knowledge about special phenomena.
An \glqq astronomer\grqq\ was usually a senior member receiving the
venerableness of the local group.
The mean age of a Maori amounted to 31 to 32 years in the
pre-European era (till $\approx$1770).
Only few attained an age of 50 \cite{orchiston}.
Although the Maori developed their own calendar taking notice of
the seasons and motion of the moon, they did not count particular
years.
It seems that they found no use in tracing an exact chronology.
The incident was remembered as it was but not the time of its
occurrence.
Durations of rulerships are not known, either.

Due to the rareness of a solar eclipse this extraordinary
experience was granted to just a few.
As an adjunct, oral tradition precludes the ability of finding a
regularity within the appearances.
Computations were not performed, nor predictions of future events
made.
Each eclipse came up entirely unexpected.
Thus, we should be very careful deploying possible accounts as
a marker for dating, since only the last \emph{real} event would
be remembered by the narrator, if he was lucky to witness it by
himself.
The reliability of any historical information passed through the
filter of oral transmission cannot be trusted for more than one
or two generations
despite being viewed as ``true'' by the culture in which they
are told.
On the other side, it would be unwise to entirely dismiss such a
traditional story, because Polynesians (as well as Aborigines in
Australia) are very gifted observers who have proven their ability
in memorisation, and also in documenting and understanding nature.
The historical core of the story can be objectively retrieved and
studied, as done by William Masse and his collaborators
\cite{snyder_2011}.
He demonstrated that the myths of Polynesians do preserve
impressively accurate details.

%
\fancyhead{}
\fancyhead[CE, CO]{\footnotesize \itshape E.\ Khalisi (2020): \titel}
\renewcommand{\headrulewidth}{0pt}


\section{Woman in the Moon}

The Maori language knows a number of terms for the sun in a
specific context.
It makes a difference when it is dealt with a reddish sun close
to the horizon, or when denoting the time of day, or when it is
used in a ceremonial act, or when personified with a deity ---
in each case a different expression was used.
Elsdon Best (1856--1931), a writer and ethnographer who interviewed
many Maori about their mythology and culture, enumerates many
personified forms of that luminary \cite{best_1922}.
Considerable respect was paid to the sun in rituals.
Best rejects some contrary opinions by early anthropologists that
no exuberant devotion would have existed throughout all of Polynesia
incorporating sacrifices and shrines, as arranged in other cultures.
The personification of a natural phenomenon will already be a
recognition of an adoration.

Eclipses of the sun and moon are such striking phenomena disturbing
normality that every culture wished for an explanation.
Any myth about eclipses proves that people spent thoughts about
their cause.
In New Zealand, the demigod Maui who takes on a variety of roles
in the folklore of many peoples throughout the entire Pacific region
is said to have fastened the sun to the moon such that, as the
former went down, the other being pulled after it \cite{yate_1835}.
At stated seasons Maui places his hand between the sun and the
earth that there will be no light.

Like in the case of the sun, various names are applied to the moon,
mostly female names.
Colloquially she is called ``Marama'', but the personified form
is ``Hina''.
Both names are widespread in Polynesia, with some dialectic changes,
adopting diverse roles in respect of the demigod Maui.
For example, Hina is sometimes made his sister, or his mother
(Hawaii), or wife (Tuamotu Islands), or daughter of other mythical
beings with varying relationship to the protagonist.
The moon itself is not considered as a deity but rather a kind of
patroness watching over labours peculiar to women such as weaving
and, most of all, childbirth.

\begin{figure}[t]
\includegraphics[width=\linewidth]{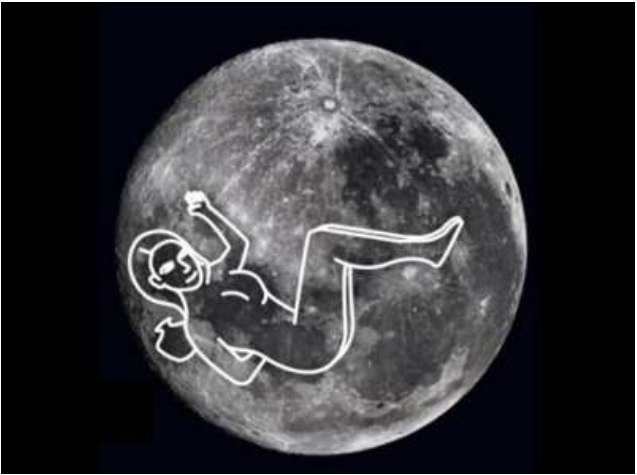}
\caption{Suggestion for Rona on the moon.}
\label{fig:rona}
\end{figure}
\enlargethispage{1ex}

In analogy to our ``Man in the Moon'' the Maori mythology renders
a ``Woman in the Moon'', her name being Rona.
She acts as a malignant being trying to attack and destroy the orb
\cite{best_1922}.
A popular fireside story is that Rona once lived on earth and went
out to a spring for water one night carrying her gourd
water-vessels.
On her way back the moon disappeared behind a cloud, and Rona
could not see the path.
She stumbled in the dark across a root protruding out of the ground
and cursed the moon for her fall.
Marama, the moon, heard this and snatched her up into the sky.
Rona tried to impede the assault by clinging to a tree, but the
tree was teared out of the ground, together with its roots, and
transferred to the moon, too.
Yet, Rona is seen there in the shape of the dark patches as well
as the tree and the gourd-vessels (Fig.\ \ref{fig:rona}).
During a lunar eclipse Rona battles with the moon, thus, the latter
cannot be seen.
After the combat the moon bathes in the ``waters of life'' and
returns reinvigorated, young and beautiful.

Similar stories about a girl who is taken to the moon is known
from a number of Siberian peoples \cite{finsternisbuch}.
In some cases they deal with an orphan, in another version she
is freezing, another again speaks of a girl with an evil stepmother.
A striking parallel to the Maori story is that the girl was
likewise on her way fetching water.


\section{Blocks of Eclipses}

Taking up a more scientific view, we were not able to detect a
firm record on a distinct eclipse for dating purposes.
In search for outstanding events in the time before the Europeans
we deployed the \textit{Five Millennium Canon of Eclipses} by
Fred Espenak  \cite{espenak}.
Table \ref{tab:nz15thcentury} presents all occurrences of magnitude
larger than 0.5 in the 14th and 15th century.

The magnitude (mag) of an eclipse denotes the ratio of the diameter
of the moon, $\theta_{\rm M}$, to the diameter of the sun,
$\theta_{\odot}$, as seen from earth:
\begin{equation}
\text{mag} = (\theta_{\odot} + \theta_{\rm M} - \Delta ) / 2 \; \theta_{\odot} ,
\end{equation}
with $\Delta$ being the distance of the centres of the two disks.
Both $\theta_{\odot}$ and $\theta_{\rm M}$ have an angular diameter
of $\approx0.5^{\circ}$, while their minute variations are due to
the alternating distance of each body at its perihelion/aphelion and
perigee/apogee, respectively.
If mag is $>$ 1, the eclipse will be total.
The magnitude is not the same as ``obscurity'' which is the fraction
of the \emph{area} of the sun's disk covered by the moon.
A solar eclipse is not necessarily noticed, since the lighting
conditions will change above a magnitude $\approx$0.75 depending
on random factors like weather, attention of the observer, and
perhaps the time of day.

\begin{table*}[t!]
\caption{Solar eclipses over New Zealand with a magnitude $>$0.50.
    The magnitude, local time and altitude refer to the today's
    capital Wellington (41.4$^{\circ}$ S, 174.8$^{\circ}$ E).
    Eclipse types: (T)otal, (A)nnular, (H)ybrid, and (P)artial.
    Clusters of special interest are highlighted.
    The lower part of the table contains only central tracks over
    the mainland in the 21st century.}
\label{tab:nz15thcentury}
%
\centering
\vspace{0.5ex}
\begin{tabular}{rlc|rrc|l}
\hline
\rowcolor{grey20}
\multicolumn{2}{c}{\cellcolor{grey20}{Date}} &
             Type & \multicolumn{1}{c}{\cellcolor{grey20}{LT}}
                           & Alt. [$^{\circ}$] & Magn. & Remark \\
\hline
1285 & Nov 28 & H & 12:39 & 68.9 & 0.770 & off the northern coast \\
1293 & Dec 29 & A & 18:06 & 15.1 & 0.867 & central in the north, sunset \\ 
\cdashline{1-7}[0.5pt/5pt]     
1301 & Aug 05 & T & 13:12 & 31.6 & 0.907 & total on the northern coast \\
1308 & Mar 23 & T & 14:05 & 37.0 & 0.519 &    \\
1310 & Jul 27 & T & 12:52 & 30.1 & 0.842 &    \\
1315 & Oct 29 & A & 13:04 & 59.4 & 0.602 &    \\
1317 & Mar 14 & T & 13:40 & 43.7 & 0.861 &    \\
1324 & Oct 19 & A &  7:56 & 32.9 & 0.520 &    \\
1330 & Jan 20 & A &  7:29 & 25.7 & 0.776 & central on the southern tip \\
1348 & Jan 31 & A & 18:11 & 11.0 & 0.531 & at sunset \\
1355 & Sep 07 & T & 12:11 & 44.6 & 0.785 &    \\
1359 & Jun 26 & A & 15:08 & 12.8 & 0.778 & central on the southern tip, sunset \\
1364 & Aug 27 & T & 11:55 & 41.8 & 0.696 &    \\
\cellcolor{grey10}
1368 & Jun 16 & A &  8:35 &  9.5 & 0.854 & central on the northern tip, sunrise \\
1369 & Nov 30 & A & 14:05 & 57.1 & 0.600 &   \\
\cellcolor{grey10}
1371 & Apr 16 & T & 13:27 & 32.3 & 0.807 &   \\
\cellcolor{grey10}
1376 & Jan 21 & T &  6:12 & 11.1 & 0.819 & off the northern coast, sunrise \\
1379 & Nov 10 & A &  8:55 & 47.5 & 0.654 &    \\
1381 & Apr 25 & P & 12:22 & 32.7 & 0.573 &    \\
1384 & Feb 21 & A &  7:15 & 16.4 & 0.832 & central on the southern tip, sunrise \\
1390 & Apr 15 & H &  9:54 & 28.7 & 0.543 &    \\
\cdashline{1-7}[0.5pt/5pt]     
1409 & Oct 09 & T & 13:04 & 52.7 & 0.986 & total on North Island \\
\cellcolor{grey10}
1424 & Jan 02 & A & 15:12 & 47.2 & 0.904 & central on southeastern coast \\
\cellcolor{grey10}
1430 & Feb 23 & T &  7:35 & 19.8 & 0.910 & total on South Island, sunrise \\
\cellcolor{grey10}
1433 & Dec 12 & A &  8:22 & 41.5 & 0.937 & central in Wellington \\
\cellcolor{grey10}
1435 & May 28 & T &  9:34 & 17.9 & 0.938 & off the SE coast, sunrise \\
1437 & Sep 30 & T & 12:38 & 53.0 & 0.730 &    \\ 
1438 & Mar 26 & A &  6:07 &$-$3.0& 0.895 & off the northern coast, sunrise \\
1440 & Jul 29 & P & 12:55 & 31.0 & 0.620 &    \\ 
1444 & May 18 & H &  7:25 &  1.3 & 0.976 & off the eastern coast, sunrise \\
1451 & Jan 03 & P & 19:09 &  4.4 & 0.640 & sunset \\
\cellcolor{grey10}
1460 & Jan 23 & A &  5:03 &$-$1.6& 0.808 & off the NE coast, sunrise \\
\cellcolor{grey10}
1463 & Nov 11 & T & 15:25 & 39.2 & 0.980 & total on South Island \\
\cellcolor{grey10}
1464 & Oct 31 & T &  5:04 &  3.5 & 0.900 & sunrise \\
1478 & Feb 03 & A & 15:25 & 40.7 & 0.647 &    \\
\cellcolor{grey10}
1484 & Mar 27 & T &  8:00 & 17.0 & 0.861 & off the southern coast, sunrise \\
\cellcolor{grey10}
1485 & Sep 09 & A & 14:40 & 32.8 & 0.855 & annular on North Island \\
\cellcolor{grey10}
1488 & Jan 14 & A &  8:05 & 33.5 & 0.745 & annular on the South Island \\
\cellcolor{grey10}
1491 & Nov 02 & T & 13:56 & 53.1 & 0.890 & total on South Island \\
\cdashline{1-7}[0.5pt/5pt]     
1507 & Jul 10 & H & 16:16 &  5.0 & 0.909 & annular on South Island, sunset \\
1516 & Jun 30 & A & 13:24 & 23.7 & 0.921 & annular on northern coast \\
\hline
2028 & Jul 22 & T & 16:02 &  8.0 & 0.854 & total on mainland, sunset \\ 
\cellcolor{grey10}
2035 & Mar 09 & A &  9:35 & 39.1 & 0.987 & central in outskirts of Wellington \\ 
\cellcolor{grey10}
2037 & Jul 13 & T & 15:25 & 11.9 & 0.955 & total on North Island, sunset \\ 
\cellcolor{grey10}
2038 & Dec 26 & T & 13:01 & 68.1 & 0.989 & total on mainland \\ 
\cellcolor{grey10}
2042 & Oct 14 & A & 15:13 & 33.8 & 0.858 & central on South Island \\ 
\cellcolor{grey10}
2045 & Feb 16 & A & 10:35 & 53.4 & 0.960 & central in Wellington \\ 
2066 & Dec 17 & T & 12:02 & 72.0 & 0.832 & total on Stewart Island \\ 
2068 & May 31 & T & 16:48 &$-$2.2& 0.866 & total on South Island, sunset \\ 
2079 & Oct 24 & A &  5:07 &  0.8 & 0.969 & central on mainland, sunrise \\ 
2096 & Nov 15 & A & 12:58 & 62.6 & 0.815 & central on North Island \\ 
(2099 & May 21 & A &  8:59 & 20.2 & 0.804 & near miss: $\approx$100 km) \\ 
\end{tabular}
\end{table*}

\begin{figure*}[t]
\centering
\includegraphics[width=\linewidth]{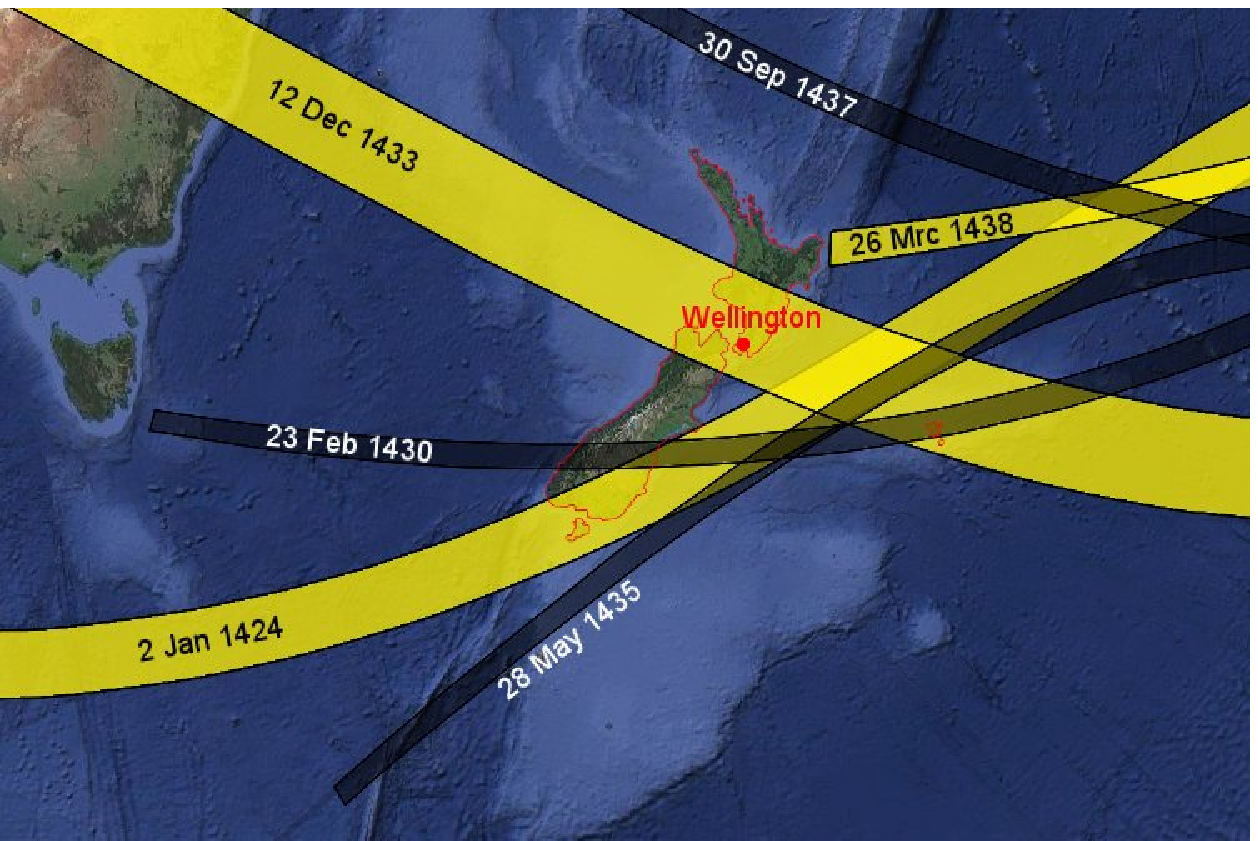}
\caption{Three annular eclipses (yellow) and three total eclipses
    (grey) within 14 years in New Zealand in the 15th century.
    The last three events of 1435, 1437, and 1438 would be seen
    partial onshore.}
\label{fig:sextett}
\end{figure*}

When peering through the list, one can detect two or three decades
harbouring ``eclipse quadruples'', meaning that four high-magnitude
events are tightly spaced.
Intriguing is the period from 1424 to 1435 when the tracks passed
over or close to the main islands of New Zealand
(Fig.\ \ref{fig:sextett}).
This interval is expandable by 3 more years accounting for two
additional partial eclipses of large magnitude ($>$0.7).
For example, the event of 1438 had a magnitude of 0.732 in
Wellington at the instant of sunrise.
There exists a quadruple point 500 km off the East coast, while on
the mainland only a small common region for just two eclipses is
discernable (1424 and 1430).
Not each of these eclipses would have been observed because of
inclement weather, but even a few within a lifetime would be a
noteworthy peculiarity.
Another quadruple within 8 years touched New Zealand soil in the
1480ies.

Remarkable is also a very tight block of three eclipses within
5 years in the 1460ies.
The first eclipse however was at sunrise when the maximum
obscuration had already passed.
The 15th century stood out, in particular, because almost each
decade offered a spectacular event, and each generation could have
experienced one or several solar eclipses, in principle.
If only half of the events in these blocks had been perceived by
the same observer, they would have left a lasting impression.
Such a state of affairs we claim, e.g., for the Egyptian pharaoh
Akhenaten who might have witnessed three high-magnitude eclipses,
see \cite{khalisi-egypt}.
One possible Maori account is discussed in the next section below.
Further triples occurred in the time intervals from 1163 to 1167
as well as 1545 to 1554 (both not in the list) with their
magnitudes ranging at somewhat lower values than above.

The second half of the 15th century was characterised by a phase
of low solar activity, the so-called Sp\"orer-Minimum.
This is a period of suspended sunspot activity lasting from 1460
to 1550 AD.
The total eclipses of 1463 and 1491 over the South Island could
have presented a pale corona and probably no prominences.
These concomitants of totality, which are not visible in partial
or annular eclipses, were already mentioned in European literature
before, but the scientific discussion picked up momentum in the
late 17th century.
An individual, who usually sees a total eclipse once in his
lifetime, has the disadvantage of no direct comparison of events.
The observer refers to the sudden darkness and its psychological
impact.
It is the multiple experience that would have raised awareness of
the existence about such things like the corona and prominences.
Ancient scholars were much too amazed about the abruptness that
they did not pay attention to such intricacies for many centuries.

\begin{figure}[t]
\includegraphics[width=\linewidth]{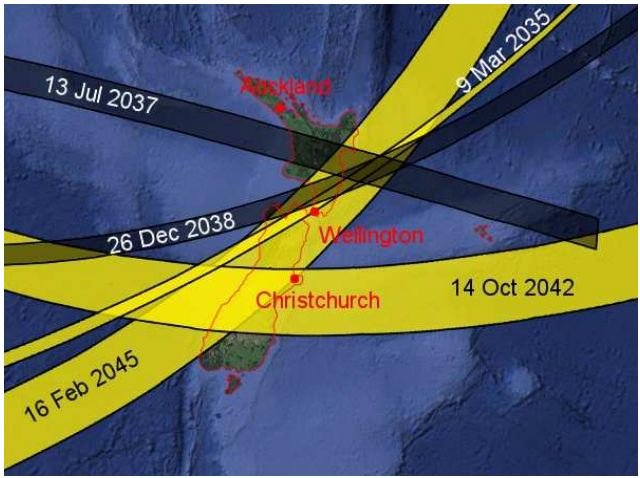}
\caption{Five eclipses between 2035 and 2045 (same colours as in
    Figure \ref{fig:sextett}).}
\label{fig:nz21cen}
\end{figure}

\begin{figure}[t]
\includegraphics[width=\linewidth]{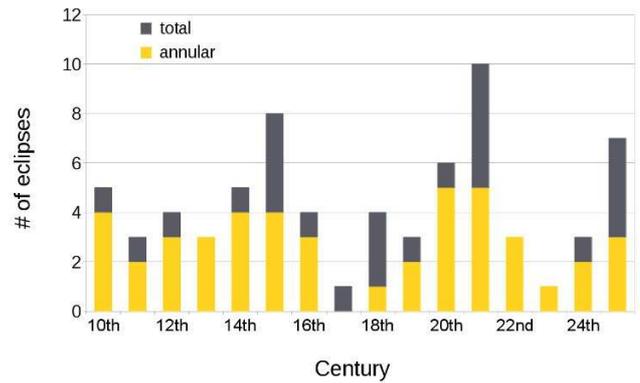}
\caption{Number central eclipses (annular + total) per century
    touching the mainland of New Zealand.}
\label{fig:eclipercen}
\end{figure}

In our time, eclipses lost their dread.
They have become a tourist attraction, and eclipse chasers will
have their fortune in the years 2035 to 2045.
New Zealand will be hit by a quintuple of central events
(Fig.\ \ref{fig:nz21cen}).
There exits a small region about 75 km off the Eastern coast of
North Island where four of the five eclipses can be seen centrally
(quadruple point).
The eclipses in the 21st century surpass by number those of the
Maori era.
The histogram in Figure \ref{fig:eclipercen} gives an overview of
central events per century that traversed the main islands during
the past millennium up until 2500 AD.
The area size of New Zealand is akin to United Kingdom, but it is
blessed with eclipses in our time.
The principal advantage of the country, however, is its elongated
shape (1400 km in North-South direction vs.\ 900 km for the UK)
boosting the probability to be crossed by an umbra or antumbra.


\section{An Ancient Eclipse Record?}

A recent study on the early Maori settlement made a dating attempt
based on an alleged eclipse from ancient times.
Ockie Simmonds and Kiyotaka Tanikawa believe that two Polynesian
tribes from the eastern Pacific arrived later than commonly
believed, in late 1408, and witnessed the solar eclipse of
9 October 1409 \cite{simmonds_2018}.
That statement made us re-check the circumstances.

The arguments rely on the genealogy of the tribes involving 24
generations of 25 years on average.
Using these numbers, the authors estimated the two seafaring groups
by tracking down the names of chieftains from their modern
ancestors at 1900 AD.
In their conclusion, that journey across the ocean would have
taken place about 600 years ago
providing an approximate time range for further analysis in more
detail.

\enlargethispage{1ex}

\begin{figure}[t]
\centering
\includegraphics[width=\linewidth]{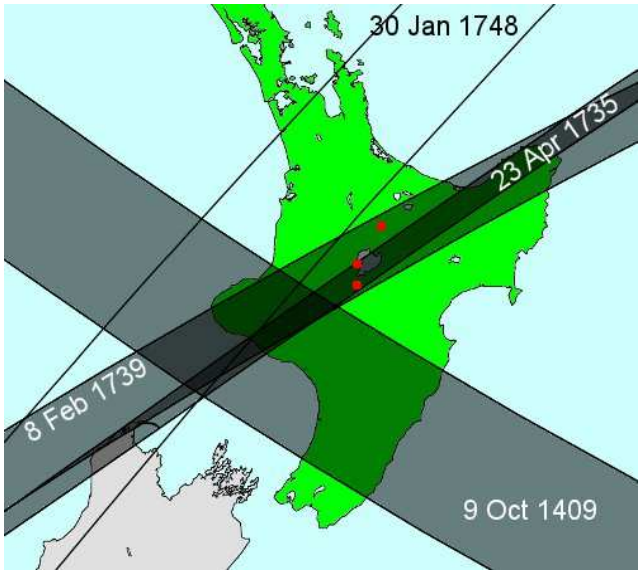}
\caption{Four total eclipses on North Island.
    Red dots show the locations of the Maori explorers in the myth
    as suggested by \cite{simmonds_2018}.
    The track of 1748 is unshaded for clarity.}
\label{fig:northisland}
\end{figure}

The mythological story to which the heroic individuals of the
tribes are connected is said to contain information on a sudden
darkening.
The persons must have witnessed the totality of an eclipse, the
authors of the study assume.
Additionally, the names of two or three geographical places on
North Island would point to the newcomers' walking tour when they
explored the volcanic plateau around Lake Taupo following their
arrival.
These places are named after the members of that tribe.
One of them, for example, is said to have been pursued by an
``evil witch''.
He sought shelter in a rock and escaped her.
That place is now sacred in the memory of the Maori and, moreover,
a government protected site.
Using extra information on key persons, the authors infer the
eclipse of 9 October 1409 as the most appropriate event that swept
over the region (Fig.\ \ref{fig:northisland}).

The work is characterised by an industrious collection of
semantical details on Maori legends, an ability of proper
recitation, and a decent sense of locality.
However, the reasoning appears weak to us from the scientific point
of view.
The authors point out in their introduction how crude and
unsatisfactory the 25-years-generation-count actually is for
establishing a chronology.
In spite of their own warnings they adhere to the relative sequence
of traditional genealogies and eventually fail to provide absolute
historical markers.
The happenings are solely pinned to that alleged eclipse.
It would be much appreciated, if a few intermediate pillars backed
those 600 years between then and now (comets, encounters of planets,
or natural disasters).
Another advantage would arise from independent evidence, for
instance, samples for radiocarbon analysis or other substantiated
material.
Even without this kind of hard evidence we cannot approve of the
smooth linearity of rulerships of 25 years, as if there were never
premature deaths nor conflicts among contender rulers nor any other
troubles affecting the timeline somehow.

Among other inconsistencies we wonder about those three protagonists
of the tribe, one of which is said to have perished in the cold.
From our point of view, this is no proof for the onset of snow,
nor does it tell anything about the season.
The authors go on suggesting locations on the North Island where
the persons might be overtaken by the shadow of the solar eclipse.
Comparing their suggested locations with the track of 1409, all of
them would have stayed outside the zone.
If accepting that sudden darkness to be an eclipse, we discovered
two other events for the North Island including Lake Taupo:
23 Apr 1735 and 8 Feb 1739 (Fig.\ \ref{fig:northisland}).
They exhibit an almost identical path.
Both would agree with the locations proposed by the authors much
better, although we express serious doubts whether their exact
positions can be retrieved at all.
There would even be a third totality in 1748, a bit outside the
region in question though.
It would be easier to suspect a slip in the story than a displacement
of the eclipse track at will in order to please the plot.
We abstain from discussing other flaws in the line of argumentation.

As pointed out by Rawiri Taonui \cite{taonui_2006}, migratory
stories contain the greatest mix of history and symbolism.
They are the most difficult to interpret and have led to much
distortion in publications as writers often overemphasised the
content.
For New Zealand more than 40 ``first arrival traditions'' are known,
and each large tribe claims one or more ancestors to have settled
there at some time between 950 and 1350.
Even in case of the lineage being true, one sole hint centred on
the eclipse will hardly suffice to withstand the test as regards
the absolute timescale.

\section{Summary}

Our investigation of solar eclipses in the southern hemisphere of
the earth revealed a good number of closely spaced tracks in the
region of New Zealand:
triples and quadruples within a decade.
Three such blocks piled up in the 15th century when the Maori
inhabited the islands.
A quintuple will commence in 2035.
Such clusterings are statistical outliers making eclipses on Earth
something special among the planets.

We gained beneficial insight into the native culture of New Zealand
that used to preserve its history in semi-mythological tales.
Being aware of the risks at judging a mental attitude completely
different from the modern European comprehension, we cannot hearken
back to substantiated accounts about an observation of an eclipse.
Astronomical dating out of myths remains precarious.
We want to encourage those being familiar with the mythology and
folklore of the Maori to search for probable eclipses but likewise
hand out advice to be cautious at interpretation.
Not every mention of a sudden darkness means an eclipse.
Sky dimming may have manifold reasons.
A total event under an overcast sky could pass unnoticed, even if
standing inside the totality zone.
There will be a lot to discover in the old narratives when
exercising due care.

\section*{Acknowledgements}

This paper is rooted in Chapter 19.4 of the Habilitation submitted
to the University of Heidelberg, Germany \cite{finsternisbuch}.
The entire work was accomplished in despite of serious hardships.
The author thanks all his friends and acquaintances who supported
him in times of trouble.
The sketch in Figure \ref{fig:rona} is by Grace Abbott published
in 2015.

\newpage


\end{document}